\documentclass[twocolumn,preprintnumbers,showpacs,amsmath,amssymb]{revtex4}
\usepackage{graphicx}
\usepackage{amssymb}

\begin{document}
\title{Security bound of cheat sensitive quantum bit commitment}
\author{Guang Ping He}
\email{hegp@mail.sysu.edu.cn}
\affiliation{School of Physics and Engineering, Sun Yat-sen University, Guangzhou 510275,
China}

\begin{abstract}
Cheat sensitive quantum bit commitment (CSQBC) loosens the security
requirement of quantum bit commitment (QBC), so that the existing
impossibility proofs of unconditionally secure QBC can be evaded. But here
we analyze the common features in all existing CSQBC protocols, and show
that in any CSQBC having these features, the receiver can always learn a
non-trivial amount of information on the sender's committed bit before it is
unveiled, while his cheating can pass the security check with a probability
not less than $50\%$. The sender's cheating is also studied. The optimal
CSQBC protocols that can minimize the sum of the cheating probabilities of
both parties are found to be trivial, as they are practically useless. We
also discuss the possibility of building a fair protocol in which both
parties can cheat with equal probabilities.%
% The general trade off between the information gained and the successful
% cheating probability is also given.
\end{abstract}

\pacs{03.67.Dd, 89.70.-a, 03.67.Mn, 03.65.Ud}
\maketitle

%\newpage

%%%%%%%%%%%%%%%%%%%%%%%%%%%%%%%%%%%%%%%%%%%%%%%%%%%%%%%%%%%%%%%%%%%%%%%%%%%%%%%%%%%%%%%%%%%%%%%%%%%%%%%%%%%%%%%%%%%%%%%%%%%%%%%%%%%%%%%%%%%%%%%%%%%%%%%%%%%%%%%%%%%%%%%%%%%%%%%%%%%%%%%%%%%%%%%%%%%%%%%%%%%%%%%%%%%%%%%%%%%%%%%%%%%%%%%%%%%%%%%%%%%%%%%%%%%%

Quantum bit commitment (QBC) is a two-party cryptography including the
following phases. In the commit phase, Alice (the sender of the commitment)
decides the value of the bit $b$ ($b=0$ or $1$) that she wants to commit,
and sends Bob (the receiver of the commitment) a piece of evidence, e.g.,
some quantum states. Later, in the unveil phase, Alice announces the value
of $b$, and Bob checks it with the evidence. The interval between the commit
and unveil phases is sometimes called the holding phase. A QBC protocol is
called unconditionally secure if any cheating can be detected with a
probability arbitrarily close to $1$. Here Alice's cheating means that she
wants to change the value of $b$ after the commit phase, while Bob's
cheating means that he tries to learn $b$ before the unveil phase.

QBC is an essential primitive for building quantum multi-party secure
computations and other \textquotedblleft post-cold-war
era\textquotedblright\ multi-party cryptographic protocols \cite{qi75,qi139}%
. Unfortunately, it is widely believed that unconditionally secure QBC is
impossible \cite{qi24,qi23}. This result, known as the Mayers-Lo-Chau (MLC)
no-go theorem, was considered as putting a serious drawback on quantum
cryptography.

To evade the problem, the concept \textquotedblleft cheat sensitive quantum
bit commitment (CSQBC)\textquotedblright\ was proposed \cite%
{qi150,qbc50,qbc78,qi197,qbc52,qbc89}, where the probability for detecting
the cheating does not need to be arbitrarily close to $1$. Instead, it
merely requires the probability to be nonzero. With this loosen security
requirement, many insecure QBC protocols can be regarded as secure CSQBC.
Therefore, at the first glance it seems that CSQBC will be very easy to
achieve.

But intriguingly, here we will show that there still exists boundary for the
security of a typical class of CSQBC. Especially, Bob can always feel free
to measure the quantum states to learn $b$, while he stands at least $50\%$
chances to escape Alice's detection.

\section*{Result}

\subsection*{Common features of CSQBC}

By checking the existing CSQBC protocols \cite%
{qi150,qbc50,qbc78,qi197,qbc52,qbc89}, we find that they all share the
following common features (note that the names Alice and Bob are used
reversely in \cite{qbc78,qbc52,qbc89}):

(1) During the holding phase, the receiver Bob owns a quantum system $\Psi $%
\ encoding Alice's committed bit $b$. ($\Psi$\ can either be prepared by the
sender Alice, or be prepared by Bob and sent to Alice, who returns it to Bob
after performing some certain operations according to her choice of $b$. It
also does not matter whether Alice prepared and kept another quantum system
entangling with $\Psi$.)

(2) Bob knows the definitions of $\rho_{0}^{B}$\ and $\rho_{1}^{B}$ directly
before the end of the commit phase. (That is, these definitions are either
clearly stated by the protocol, or announced to Bob\ by Alice classically.
Bob does not need to perform operations on any quantum system to gain
knowledge of these definitions.) Here $\rho_{0}^{B}$\ and $\rho_{1}^{B}$\
are the density matrices of Bob's $\Psi$ corresponding to $b=0$ and $b=1$,
respectively.

(3) To detect Bob's cheating, at the unveil phase Alice can check whether
the state of $\Psi$\ is intact. (It does not matter whether the entire $\Psi$%
\ or only a small part can be checked.)

(4) To detect Alice's cheating, at the unveil phase Bob can learn a
nontrivial amount of information on the value of $b$ from $\Psi$, even
without any help from Alice.

The last feature indicates that there exists at least one operation known to
Bob, which can output a bit $b^{\prime }$ when being applied on $\Psi $, and
$b^{\prime }=b$ should occur with a probability larger than $1/2$. As a
result, there must be $\rho _{0}^{B}\neq \rho _{1}^{B}$. This is a main
difference from the original QBC, where there is generally $\rho
_{0}^{B}\simeq \rho _{1}^{B}$ so that it can be unconditionally secure
against dishonest-Bob. %{\large Otherwise,
%i.e., if }$\rho_{0}^{B}=\rho_{1}^{B}${\large , then the protocol will return
%to the original QBC studied in the MLC no-go theorem,\ instead of CSQBC. In
%this case, as Bob always knows nothing about }$b${\large without the help from
%Alice, he cannot check whether the commitment is honestly made until Alice
%provided him with additional informations (including quantum states) in the
%unveil phase. Then dishonest-Alice is surely prone to provide misleading
%information to Bob if she wants to alter her committed value, making the
%protocol insecure against Alice's cheating.}

The original purpose of CSQBC having these features is as follows. Alice's
cheating strategy suggested in the MLC no-go theorem is based on the
Hughston-Jozsa-Wootters (HJW) theorem \cite{qi73}, which applies to the case
$\rho _{0}^{B}\simeq \rho _{1}^{B}$. Therefore with feature (4), i.e., $\rho
_{0}^{B}\neq \rho _{1}^{B}$, Alice's cheating becomes detectable so that\
the MLC no-go theorem can be evaded. On the other hand, if Bob takes
advantages of $\rho _{0}^{B}\neq \rho _{1}^{B}$\ and performs measurements
to discriminate the committed bit $b$, the quantum state will be disturbed.
In this case, with feature (3) Bob's cheating will be detected with a
certain probability when Alice asks\ him to return the quantum state and
checks wether it remains undisturbed, so that the goal of CSQBC can be met.

But with a rigorous quantitative analysis on the probability of detecting
Bob's cheating, we will find that it is always not sufficiently large when
Bob applies some specific measurements. Therefore any CSQBC protocol having
the above four features will be bounded by the security limit below.

\subsection*{Notations and Bob's cheating strategy}

According to Eq. (9.22) of \cite{qi366}, the trace distance $D(\rho
_{0}^{B},\rho _{1}^{B})\equiv tr\left\vert \rho _{0}^{B}-\rho
_{1}^{B}\right\vert /2$\ (where $\left\vert A\right\vert \equiv \sqrt{%
A^{\dagger }A}$) between $\rho _{0}^{B}$ and $\rho _{1}^{B}$\ satisfies%
\begin{equation}
D(\rho _{0}^{B},\rho _{1}^{B})=\max_{P}tr(P(\rho _{0}^{B}-\rho _{1}^{B})),
\end{equation}%
where the maximization is taken over all positive operators $P\leq I$, with $%
I$ being the identity operator. The above feature (2) of CSQBC guarantees
that Bob knows how $\rho _{0}^{B}$\ and $\rho _{1}^{B}$ are defined. Thus he
can find the positive projectors $P=P_{m}$\ that maximizes $tr(P(\rho
_{0}^{B}-\rho _{1}^{B}))$. If $\rho _{0}^{B}$\ stands a higher probability
to be projected successfully than $\rho _{1}^{B}$\ when applying $P_{m}$,
then we takes $P_{0}\equiv P_{m}$\ and $P_{1}\equiv I-P_{m}$. Otherwise we
takes $P_{0}\equiv I-P_{m}$\ and $P_{1}\equiv P_{m}$. Feature (1) ensures
that Bob owns the system $\Psi $\ encoding Alice's committed bit $b$ during
the holding phase. Therefore, by applying the positive operator-valued
measure (POVM) $\{P_{0}^{\dag }P_{0},P_{1}^{\dag }P_{1}\}$\ on $\Psi $, Bob
can discriminate between $\rho _{0}^{B}$\ and $\rho _{1}^{B}$ and learn
Alice's committed $b$ with the maximal probability allowed by $D(\rho
_{0}^{B},\rho _{1}^{B})$.

To analyze rigorously the probability for Bob to escape Alice's detection
with this POVM, let $H$\ be the global Hilbert space constructed by all
possible states of $\Psi $ (either $b=0$ or $1$). Since $P_{0}$, $P_{1}$\
are positive projectors, there exists an orthonormal basis $\{\left\vert
e_{i}\right\rangle \}$ of $H$ (the following proof remains valid regardless
whether $\{\left\vert e_{i}\right\rangle \}$ is known to Alice or Bob), in
which $P_{0}$, $P_{1}$ can be expressed as%
\begin{eqnarray}
P_{0} &=&\sum\limits_{i}\left\vert e_{i}^{(0)}\right\rangle \left\langle
e_{i}^{(0)}\right\vert ,  \nonumber \\
P_{1} &=&\sum\limits_{i}\left\vert e_{i}^{(1)}\right\rangle \left\langle
e_{i}^{(1)}\right\vert ,  \label{projector2}
\end{eqnarray}%
where $\{\left\vert e_{i}^{(0)}\right\rangle \}\cup \{\left\vert
e_{i}^{(1)}\right\rangle \}=\{\left\vert e_{i}\right\rangle \}$.

Meanwhile, before Bob applying any measurement, the general form of the
initial state of $\Psi $\ can always be written as%
\begin{align}
\left\vert \Phi \otimes \Psi \right\rangle _{ini}& =\sqrt{\alpha }%
\sum\limits_{i}\lambda _{i}^{(0)}\left\vert f_{i}^{(0)}\right\rangle \otimes
\left\vert e_{i}^{(0)}\right\rangle   \nonumber \\
& +\sqrt{\beta }\sum\limits_{i}\lambda _{i}^{(1)}\left\vert
f_{i}^{(1)}\right\rangle \otimes \left\vert e_{i}^{(1)}\right\rangle ,
\label{initial}
\end{align}%
where $0\leq \alpha \leq 1$, $\beta =1-\alpha $, and $\sum\limits_{i}\left%
\vert \lambda _{i}^{(0)}\right\vert ^{2}=\sum\limits_{i}\left\vert \lambda
_{i}^{(1)}\right\vert ^{2}=1$\ (sum over all possible $i$ within each
corresponding subspace). The values of $\alpha $, $\beta $, $\lambda
_{i}^{(0)}$'s and $\lambda _{i}^{(1)}$'s\ are chosen by Alice according to
the value of her committed bit $b$.\ Here $\Phi $ is a quantum system that
Alice may introduce and keep to herself, which entangles with Bob's $\Psi $.
All $\left\vert f_{i}^{(0)}\right\rangle $'s and $\left\vert
f_{i}^{(1)}\right\rangle $'s are the vectors describing the state of $\Phi $%
, which are not required to be orthogonal to each other. In the case where
Alice does not introduce such a system, we can simply set all $\left\vert
f_{i}^{(0)}\right\rangle $'s and $\left\vert f_{i}^{(1)}\right\rangle $'s to
be equal, so that Eq. (\ref{initial}) still applies.

\subsection*{The security bound on Bob's cheating}

As elaborated in the 1st subsection of Methods section, when dishonest-Bob
applies the above POVM $\{P_{0}^{\dag }P_{0},P_{1}^{\dag }P_{1}\}$ on $\Psi $%
, we find that the probability for Bob's cheating to pass Alice's detection
successfully is%
\begin{equation}
P_{B}=\frac{1}{2}+\frac{1}{2}(2\alpha -1)^{2},  \label{pc}
\end{equation}%
and the amount of mutual information he obtained is%
\begin{equation}
I_{m}=1-h(\alpha ).  \label{Im}
\end{equation}%
Here $h(\alpha )\equiv -\alpha \log _{2}\alpha -(1-\alpha )\log
_{2}(1-\alpha )$\ is the binary entropy function.

With Eqs. (\ref{pc}) and (\ref{Im}), we plot $P_{B}$ and $I_{m}$\ as a
function of $\alpha $\ in FIG. 1. Since $0\leq \alpha \leq 1$, FIG. 1 and
Eq. (\ref{pc}) both gives%
\begin{equation}
P_{B}\geq 50\%.  \label{bound}
\end{equation}%
The minimum $P_{B}=50\%$\ can be reached when Alice chooses $\alpha =0.5$.
Thus we come to the conclusion that Bob can always learn Alice's committed $%
b $ with the maximal probability allowed by the trace distance between $\rho
_{0}^{B}$ and $\rho _{1}^{B}$, while his cheating stands at least $50\%$
chance to escape Alice's detection.

\begin{figure}[tbp]
\includegraphics{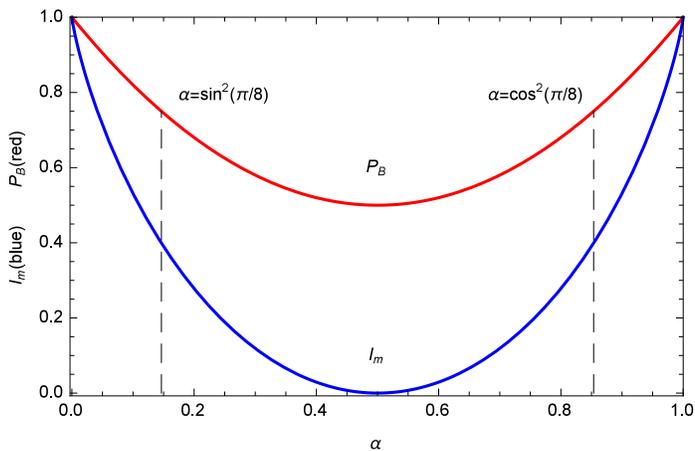}
\caption{Bob's successful cheating probability $P_{B}$ (red line) and mutual
information $I_{m}$ (blue line) on Alice's committed bit $b$ as a function
of $\protect\alpha $ that Alice chooses for the initial state Eq. (3). The
dash lines mark the values for the protocol in Ref. [5].}
\label{fig:epsart}
\end{figure}

It may look weird that FIG. 1 seems to indicate that the more amount of
information that Bob obtains, the easier he can pass Alice's detection. But
we must note that the amount of Bob's information is not chosen by himself.
Instead, it is determined by the value of $\alpha $ that Alice chooses. That
is, once Alice determines which state is used for encoding her committed
bit, the maximum amount of information that Bob can obtain is also fixed.

On the other hand, the above result indicates that Alice should make $\alpha
$\ as close to $0.5$\ as possible, so that Bob's information and successful
cheating probability can be minimized. However, note that she has to choose
the initial state Eq. (\ref{initial}) within the range restricted by the
protocol. Due to the feature (4) of CSQBC, the trace distance $D(\rho
_{0}^{B},\rho _{1}^{B})$\ has to be nonzero, Therefore, generally $\alpha $\
cannot be made very close to $0.5$, as we will see in the examples below.

\subsection*{Examples}

In the CSQBC protocol in \cite{qi150}, Bob's system $\Psi $\ is a single
qubit, whose state is either $\left\vert 0\right\rangle $\ or $\left\vert
-\right\rangle $\ ($\left\vert 1\right\rangle $\ or $\left\vert
+\right\rangle $) when Alice commits $b=0$ ($b=1$). Here $\left\vert
0\right\rangle $\ and $\left\vert 1\right\rangle $\ are orthogonal to each
other, $\left\vert \pm \right\rangle \equiv (\left\vert 0\right\rangle \pm
\left\vert 1\right\rangle )/\sqrt{2}$. So we have $\rho _{0}^{B}=(\left\vert
0\right\rangle \left\langle 0\right\vert +$\ $\left\vert -\right\rangle
\left\langle -\right\vert )/2$ and $\rho _{1}^{B}=(\left\vert 1\right\rangle
\left\langle 1\right\vert +$\ $\left\vert +\right\rangle \left\langle
+\right\vert )/2$. Define%
\begin{eqnarray}
\left\vert e^{(0)}\right\rangle &\equiv &\cos (-\pi /8)\left\vert
0\right\rangle +\sin (-\pi /8)\left\vert 1\right\rangle ,  \nonumber \\
\left\vert e^{(1)}\right\rangle &\equiv &\cos (3\pi /8)\left\vert
0\right\rangle +\sin (3\pi /8)\left\vert 1\right\rangle .
\end{eqnarray}%
Then Bob's operation for maximally discriminating $\rho _{0}^{B}$\ and $\rho
_{1}^{B}$\ is to measure $\Psi $\ in the basis $\{\left\vert
e^{(0)}\right\rangle ,\left\vert e^{(1)}\right\rangle \}$, i.e., he applies
the projector $P_{0}=\left\vert e^{(0)}\right\rangle \left\langle
e^{(0)}\right\vert $. When the projection is successful (unsuccessful), he
takes $b^{\prime }=0$ ($b^{\prime }=1$) as the decoded result. With this
method, $b^{\prime }$\ will match Alice's actual committed bit $b$ with the
probability $\cos ^{2}(\pi /8)\simeq 85.36\%$. Meanwhile, Alice's four input
states can be expanded in the $\{\left\vert e^{(0)}\right\rangle ,\left\vert
e^{(1)}\right\rangle \}$\ basis as%
\begin{align}
\left\vert 0\right\rangle & =\cos (\pi /8)\left\vert e^{(0)}\right\rangle
+\sin (\pi /8)\left\vert e^{(1)}\right\rangle ,  \nonumber \\
\left\vert -\right\rangle & =\cos (\pi /8)\left\vert e^{(0)}\right\rangle
-\sin (\pi /8)\left\vert e^{(1)}\right\rangle ,  \nonumber \\
\left\vert 1\right\rangle & =-\sin (\pi /8)\left\vert e^{(0)}\right\rangle
+\cos (\pi /8)\left\vert e^{(1)}\right\rangle ,  \nonumber \\
\left\vert +\right\rangle & =\sin (\pi /8)\left\vert e^{(0)}\right\rangle
+\cos (\pi /8)\left\vert e^{(1)}\right\rangle .
\end{align}%
Comparing with Eq. (\ref{initial}), we can see that there is either $\alpha
=\cos ^{2}(\pi /8)$\ or $\alpha =\sin ^{2}(\pi /8)$. Substitute them into
Eq. (\ref{pc}) will both yield $P_{B}=\sin ^{4}(\pi /8)+\cos ^{4}(\pi
/8)=75\%$. That is, in the CSQBC protocol in \cite{qi150}, Bob can learn
Alice's committed bit with reliability $85.36\%$\ (i.e., his mutual
information is $1-h(0.8536)\simeq 0.4$ bit) before the unveil phase, while
he can pass Alice's security check with probability $75\%$. This protocol is
corresponding to the dash lines in our FIG. 1.

Another example can be found in \cite{HeComment}, where we illustrated how
our above cheating strategy applies on the CSQBC protocol in \cite{qbc52}.
This protocol looks more complicated than the one in \cite{qi150}, as the
committed bit $b$ is encoded with many qubits, instead of a single one. The
authors of \cite{qbc52} merely analyzed the individual attack of the
receiver (note that they used the names Alice and Bob reversely) where the
qubits are measured one by one. Then it is concluded that the cheating can
be detected with a probability arbitrarily close to $1$. But as we shown
above, instead of individual measurements, the dishonest receiver can apply
a two-element POVM $\{P_{0}^{\dag }P_{0},P_{1}^{\dag }P_{1}\}$ on the entire
state encoding the committed bit. When this state consists of many qubits,
each basis vector $\left\vert e_{i}\right\rangle $ of the Hilbert space $H$
is a multi-level state describing all qubits. Thus the projectors $P_{0}$, $%
P_{1}$ in Eq. (\ref{projector2}) are actually collective measurements. The
detailed form of $P_{0}$, $P_{1}$ is given in Eq. (2) of \cite{HeComment}.
As a result, it was further elaborated there that this collective
measurement is as effective as individual measurements on learning the
committed bit, while it causes much less disturbance on the multi-qubit
state. Once again, the probability for the cheater to escape the detection
was shown \cite{HeComment} to be not less than $50\%$. With the increase of
the qubit number $n$, this probability can even be arbitrarily close to $%
100\%$.

%\cite{qi150,qbc50,qi197} presented their security level correctly, while Refs.
%\cite{qbc52,qbc89} (and \cite{qbc78}?) missed to notice the existence of the
%kind of attack like what we described above.

\subsection*{Alice's cheating strategy}

Alice's cheating strategy used in the MLC no-go theorem requires the
condition $\rho _{0}^{B}\simeq \rho _{1}^{B}$, which no longer holds in
CSQBC. Nevertheless, she can still apply the same strategy in CSQBC and try
her luck. To give a detailed description of the strategy, first let us model
the coding method in CSQBC more precisely. For generality, consider that in
the protocol, besides Bob's system $\Psi $, there is another system $E$.
Alice's different committed values of $b$ is encoded with different states
of the combined system $E\otimes \Psi $. System $E$ is kept at Alice's side
during the commit and holding phases, and is required to be sent to Bob at
the unveil phase to justify Alice's commitment. Let $\rho _{0}^{EB}$\ and
$\rho _{1}^{EB}$\ denote the density matrices of $E\otimes \Psi $
corresponding to $b=0$ and $b=1$, respectively. Note that in all existing
CSQBC protocols \cite{qi150,qbc50,qbc78,qi197,qbc52,qbc89}, there is no such
a system $E$. But we include it here, so that the model can cover more
protocols that may be proposed in the future.

In this scenario, Alice's cheating strategy is as follows. At the beginning
of the protocol she introduces an ancillary system $\Phi $ which is a copy of
$E\otimes \Psi $. Since the fidelity $F(\rho _{0}^{EB},\rho _{1}^{EB})\equiv
tr\sqrt{(\rho _{0}^{EB})^{1/2}\rho _{1}^{EB}(\rho _{0}^{EB})^{1/2}}$ between
$\rho _{0}^{EB}$\ and $\rho _{1}^{EB}$\ satisfies \cite{qi366}%
\begin{equation}
F(\rho _{0}^{EB},\rho _{1}^{EB})=\max_{\left\vert \psi _{0}\right\rangle
,\left\vert \psi _{1}\right\rangle }\left\vert \left\langle \psi _{0}\right.
\left\vert \psi _{1}\right\rangle \right\vert ,
\end{equation}%
where the maximization is over all purifications $\left\vert \varphi
_{0}\right\rangle $ of $\rho _{0}^{EB}$ and $\left\vert \varphi
_{1}\right\rangle $\ of $\rho _{1}^{EB}$ into $\Phi \otimes E\otimes \Psi $,
Alice finds the real and positive $\left\vert \psi _{0}\right\rangle $, $%
\left\vert \psi _{1}\right\rangle $ that reach the maximum, i.e.,%
\begin{equation}
F(\rho _{0}^{EB},\rho _{1}^{EB})=\left\langle \psi _{0}\right. \left\vert
\psi _{1}\right\rangle =\left\langle \psi _{1}\right. \left\vert \psi
_{0}\right\rangle .
\end{equation}%
Then she prepares the initial state of $\Phi \otimes E\otimes \Psi $ as%
\begin{equation}
\left\vert \psi _{c}\right\rangle =\frac{\left\vert \psi _{0}\right\rangle
+\left\vert \psi _{1}\right\rangle }{N},  \label{psaic}
\end{equation}%
where the normalization constant%
\begin{equation}
N=\sqrt{2+\left\langle \psi _{0}\right. \left\vert \psi _{1}\right\rangle
+\left\langle \psi _{1}\right. \left\vert \psi _{0}\right\rangle }.
\end{equation}%
She uses this state to complete the rest of the commit protocol. With this
method, the value of $b$ is not determined during the commit phase.

In the unveil phase, Alice decides whether she wants to unveil $b=0$ or $b=1$%
. Then she simply uses $\left\vert \psi _{c}\right\rangle $\ as $\left\vert
\psi _{b}\right\rangle $ to complete the protocol. From the symmetry of $%
\left\vert \varphi _{0}\right\rangle $ and $\left\vert \varphi
_{1}\right\rangle $\ in Eq.(\ref{psaic}), we can see that her successful
cheating probabilities for $b=0$ and $b=1$ are both%
\begin{eqnarray}
P_{A} &=&\left\vert \left\langle \psi _{0}\right. \left\vert \psi
_{c}\right\rangle \right\vert ^{2}=\frac{(1+\left\langle \psi _{0}\right.
\left\vert \psi _{1}\right\rangle )(1+\left\langle \psi _{1}\right.
\left\vert \psi _{0}\right\rangle )}{2+\left\langle \psi _{0}\right.
\left\vert \psi _{1}\right\rangle +\left\langle \psi _{1}\right. \left\vert
\psi _{0}\right\rangle }  \nonumber \\
&=&\frac{1+F(\rho _{0}^{EB},\rho _{1}^{EB})}{2}.  \label{pa}
\end{eqnarray}%
Therefore, in any specific CSQBC protocol, the Alice's exact cheating
probability can be calculated once the definition of $\rho _{0}^{EB}$, $\rho
_{1}^{EB}$ is known.

\subsection*{The optimal protocols are trivial}

Now we will try to find the CSQBC protocols which can optimally detect the
cheating of both parties, i.e., minimizing the sum of Alice's and Bob's
cheating probabilities.

Note that Eq. (\ref{pc}) depends on the specific value of $\alpha $ in the
state Eq. (\ref{initial}) that Alice chooses in a single run of the
protocol, while $F(\rho _{0}^{EB},\rho _{1}^{EB})$ in Eq. (\ref{pa}) is the
statistical result of all the legitimate states allowed by the protocol. Thus
it is hard to compare Eq. (\ref{pa}) and Eq. (\ref{pc}) directly and give a
general result without knowing the details on the composition of $\rho
_{b}^{EB}$ in a specific protocol.

Fortunately, in all existing CSQBC protocols \cite%
{qi150,qbc50,qbc78,qi197,qbc52,qbc89}, there is no system $E$. The form of
the states of Bob's system $\Psi $ alone carries all the information of $b$.
Thus the trace distance $D(\rho _{0}^{EB},\rho _{1}^{EB})=D(\rho
_{0}^{B},\rho _{1}^{B})$. For any protocol of this kind (as well as
protocols having system $E$ but still satisfying $D(\rho _{0}^{EB},\rho
_{1}^{EB})=D(\rho _{0}^{B},\rho _{1}^{B})$), we can replace both $\alpha $\
and $F(\rho _{0}^{EB},\rho _{1}^{EB})$\ with $D(\rho _{0}^{B},\rho _{1}^{B})$%
, as elaborated in the 2nd subsection of Method, where we obtain

\begin{equation}
P_{A}\geq 1-\frac{D(\rho _{0}^{B},\rho _{1}^{B})}{2},  \label{pa lower}
\end{equation}%
and%
\begin{equation}
P_{B}\geq \frac{1+D(\rho _{0}^{B},\rho _{1}^{B})^{2}}{2}.  \label{pb lower}
\end{equation}

These two equations suggest that $P_{A}$\ and\ $P_{B}$\ cannot be minimized
simultaneously in the same protocol, because reducing $P_{A}$\ requires a
higher $D(\rho _{0}^{B},\rho _{1}^{B})$, while it will result in a higher $%
P_{B}$ at the same time.

Moreover, we must note that the above $P_{A}$\ and\ $P_{B}$\ are obtained
assuming that the actions of both parties in the protocol will always be
checked. But this is impossible, because they share the same system $\Phi
\otimes E\otimes \Psi $. In the unveil phase, either Bob will measure $%
E\otimes \Psi $ to check Alice's action, or he is required to return $%
\Psi $ to Alice who checks his action. These cannot be done simultaneously.
Suppose that in a CSQBC protocol, Bob's action is checked with probability $%
\zeta $ ($0\leq \zeta \leq 1$), and Alice's action is checked with
probability $1-\zeta $. When one's action is not checked, he/she can cheat
successfully with probability $1$. Thus the cheating probabilities $P_{A}$\
and\ $P_{B}$\ should be replaced by%
\begin{equation}
P_{A}^{\ast }=\zeta +(1-\zeta )P_{A}  \label{pas}
\end{equation}%
and%
\begin{equation}
P_{B}^{\ast }=(1-\zeta )+\zeta P_{B},  \label{pbs}
\end{equation}%
respectively. Combining them with Eqs. (\ref{pa lower}) and (\ref{pb lower}%
), we find

\begin{eqnarray}
P_{A}^{\ast }+P_{B}^{\ast } &\geq &2-\frac{\zeta +D(\rho _{0}^{B},\rho
_{1}^{B})}{2}  \nonumber \\
&&+\zeta D(\rho _{0}^{B},\rho _{1}^{B})\frac{1+D(\rho _{0}^{B},\rho _{1}^{B})%
}{2}.  \label{psum}
\end{eqnarray}%
Since $0\leq \zeta \leq 1$\ and $0\leq D(\rho _{0}^{B},\rho _{1}^{B})\leq 1$%
, we find another security lower bound of CSQBC%
\begin{equation}
P_{A}^{\ast }+P_{B}^{\ast }\geq \frac{3}{2}.  \label{bound2}
\end{equation}

\begin{figure}[tbp]
\includegraphics{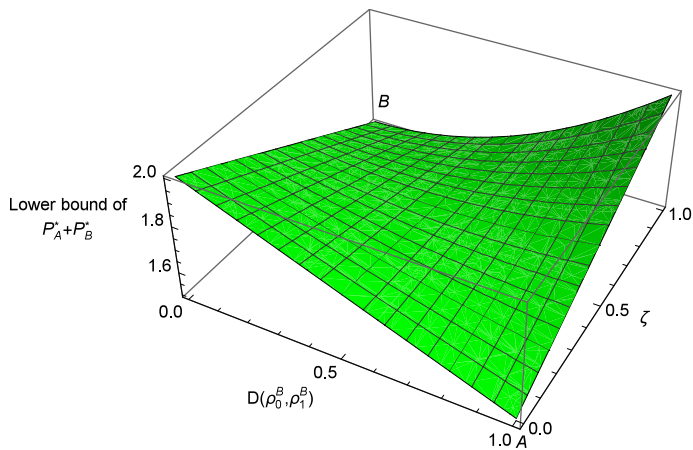}
\caption{The lower bound of the sum of the cheating probabilities $P_{A}^{\ast }+P_{B}^{\ast }$\
as a function of the trace distance $D(\rho _{0}^{B},\rho _{1}^{B})$ and the
probability $\zeta $ with which Bob's action is checked. $A$ and $B$ denote
the points that reach the minimum $P_{A}^{\ast }+P_{B}^{\ast }=3/2$.}
\label{fig:epsart}
\end{figure}

To find the optimal protocol that can reach this bound, we plot the lower
bound of $P_{A}^{\ast
}+P_{B}^{\ast }$\ as a function of $D(\rho _{0}^{B},\rho _{1}^{B})$ and $%
\zeta $\ in FIG. 2 according to Eq. (\ref{psum}). It shows that there are two
types of protocols that can both reach the minimum $P_{A}^{\ast
}+P_{B}^{\ast }=3/2$, denoted as points $A$ and $B$ in FIG. 2, respectively,
with the parameters (A) $D(\rho _{0}^{B},\rho _{1}^{B})=1$, $\zeta =0$,
and (B) $D(\rho _{0}^{B},\rho _{1}^{B})=0$, $\zeta =1$. Type (A) protocols
mean that $\rho _{0}^{B}$ and $\rho _{1}^{B}$ are orthogonal so that $%
P_{A}^{\ast }$ reaches its minimum $1/2$.
% In this case $P_{A}^{\ast }|_{b=0}+P_{B}^{\ast }|_{b=1}=1$, where $%
%P_{A}^{\ast }|_{b=0}$ ($P_{B}^{\ast }|_{b=1}$)\ is Alice's successful
%probability for unveiling $b=0$ ($b=1$). Consequently, according to the
%definition in \cite{qi581}, this should not be regarded as a successful
%cheating. On the other hand,
However, $\rho _{0}^{B}$ and $\rho _{1}^{B}$ can be distinguished
perfectly and Bob's action is never checked. Thus $P_{B}^{\ast }=1$,
i.e., he can always learn Alice's committed $b$ with reliability $1$ and
never get caught. In type (B) protocols, $\rho _{0}^{B}=\rho _{1}^{B}$ so
that Bob learns nothing about $b$. But Alice's action is never checked so
that she can unveil $b$ as whatever she wants, with a successful cheating
probability $P_{A}^{\ast }=1$. Therefore, we can see that these optimal
protocols are all trivial as they are completely insecure against one of the
parties. Thus they do not seem to have any practical usage.

\subsection*{The fair protocol}

Since the protocols that can minimize $P_{A}^{\ast }+P_{B}^{\ast }$\ all
look useless, let us consider the protocol satisfying $P_{A}^{\ast
}=P_{B}^{\ast }$ so that it is fair for both parties, and try to minimize $%
P_{A}^{\ast }$, $P_{B}^{\ast }$ in this case. From Eq. (\ref{37}) we can see
that the inequality Eq. (\ref{pb
lower}) can become equality when $\bar{\alpha}^{2}=\overline{\alpha^{2}}$, i.e.,
all the states allowed to be chosen in the protocol for committing the same $%
b$ value should have the same $\alpha $ value. Also, note that the lowest
bounds in Eqs. (\ref{pa lower}) and (\ref{psum}) cannot be reached by most $%
D(\rho _{0}^{B},\rho _{1}^{B})$, because these inequalities can become
equalities if and only if $F(\rho _{0}^{B},\rho _{1}^{B})=1-D(\rho
_{0}^{B},\rho _{1}^{B})$, which requires $\rho _{0}^{B}=\rho _{1}^{B}$.
Therefore, only the above optimal protocols can reach these bound. For this
reason, to calculate $P_{A}^{\ast }$ precisely in\ other protocols, we
should use Eq. (\ref{pa}) instead of Eq. (\ref{pa lower}). To compute $%
F(\rho _{0}^{EB},\rho _{1}^{EB})$\ in Eq. (\ref{pa}), for simplicity we consider
only the protocols in which there are%
\begin{eqnarray}
\rho _{0}^{EB} &=&\rho _{0}^{B}=\left[
\begin{array}{cc}
\alpha & 0 \\
0 & 1-\alpha%
\end{array}%
\right] ,  \nonumber \\
\rho _{1}^{EB} &=&\rho _{1}^{B}=\left[
\begin{array}{cc}
1-\alpha & 0 \\
0 & \alpha%
\end{array}%
\right] ,
\end{eqnarray}%
then%
\begin{eqnarray}
F(\rho _{0}^{B},\rho _{1}^{B}) &=&2\sqrt{\alpha (1-\alpha )},  \nonumber \\
D(\rho _{0}^{B},\rho _{1}^{B}) &=&2\alpha -1.
\end{eqnarray}%
Combining them with Eqs. (\ref{pa}), (\ref{pas}), (\ref{pbs}) and (\ref{pb
lower}) (the latter becomes equality once we choose $\bar{\alpha}^{2}=%
\overline{\alpha^{2}}$), then by solving $P_{A}^{\ast }=P_{B}^{\ast }$ we yield%
\begin{equation}
\zeta =\frac{2\sqrt{\alpha (1-\alpha )}-1}{(2\alpha -1)^{2}+2\sqrt{\alpha
(1-\alpha )}-2}.
\end{equation}%
Any protocol satisfying this equation is fair for both parties. Now let us
find the minimal value of $P_{A}^{\ast }=P_{B}^{\ast }$. Substituting this $%
\zeta $ into Eq. (\ref{pbs}), we obtain%
\begin{equation}
P_{A}^{\ast }=P_{B}^{\ast }=\frac{(2\sqrt{\alpha (1-\alpha )}+1)(2\alpha
^{2}-2\alpha +1)-2}{4\alpha ^{2}-4\alpha +2\sqrt{\alpha (1-\alpha )}-1}.
\label{pasfair3}
\end{equation}%
By solving $dP_{A}^{\ast }/d\alpha =0$, we find that the minimal cheating
probabilities in such protocols are $P_{A}^{\ast }=P_{B}^{\ast }=0.904$,
which can be obtained when $\alpha \simeq 0.885$, i.e., $\sqrt{\alpha }%
\simeq 0.941\simeq \cos (19.85^{\circ })$. In this case $\zeta \simeq 0.469$.

A simple protocol having these parameters is: Alice sends Bob the state $%
\cos (19.85^{\circ })\left\vert 0\right\rangle \pm \sin (19.85^{\circ
})\left\vert 1\right\rangle $\ ($\sin (19.85^{\circ })\left\vert
0\right\rangle \pm \cos (19.85^{\circ })\left\vert 1\right\rangle $) if she
wants to commit $b=0$\ ($b=1$). In the unveil phase, with probability $\zeta
\simeq 0.469$ Bob returns the state and Alice checks whether it remains
undisturbed, with probability $1-\zeta \simeq 0.531$ Bob measures the state
and checks whether it agrees with the value of Alice unveiled $b$.

Nevertheless, there is the difficulty in finding a method for deciding which
party will be checked in a single run of the protocol. Dishonest Alice (Bob)
would like to decrease $1-\zeta $\ ($\zeta $) so that $P_{A}^{\ast }$ ($%
P_{B}^{\ast }$)\ can be raised. Thus they do not trust each other and may
not collaborate. The CSQBC protocol in \cite{qi150} adopts a process called
\textquotedblleft the game\textquotedblright\ to handle this problem, which
is very similar to quantum coin flipping (QCF) protocols \cite{qi365}.
However, Ishizaka \cite{qi442v3} showed that this process provides extra
security loophole to Bob, so that there is a cheating strategy for him to
learn $b$ with reliability $61.79\%$\ (which is lower than what can be
obtained with our cheating strategy, as calculated in the Examples section)
while passing Alice's check with probability $100\%$ (which is higher than
that of our strategy). It was further shown in \cite{qi442} that due to the
inexistence of ideal black-boxed QCF, any CSQBC protocol based on biased QCF
cannot be secure. Therefore, it remains unclear how to build a fair CSQBC
protocol with $P_{A}^{\ast }=P_{B}^{\ast }$ while minimizing $P_{A}^{\ast }$
and $P_{B}^{\ast }$.

\section*{Discussion}

In summary, we showed that any CSQBC protocol having the above four features
is subjected to the security bound Eq. (\ref{bound}). Protocols satisfying $%
D(\rho _{0}^{EB},\rho _{1}^{EB})=D(\rho _{0}^{B},\rho _{1}^{B})$ is further
bounded by Eq. (\ref{bound2}). Note that the insecurity of QCF-based CSQBC
protocols (e.g., \cite{qi150,qbc50}) was already pinpointed out in \cite%
{qi442v3,qi442}. But our proof also applies to the non-QCF-based ones.

Our result should not be simply considered as a generalization of the MLC
no-go proof. Instead, it is a complement. This is because the MLC no-go
proof applies to QBC protocol with $\rho _{0}^{B}\simeq \rho _{1}^{B}$. But
as pointed out in \cite{qbc52}, CSQBC does not need to satisfy this
requirement so that it may evade the MLC theorem. On the contrary, our proof
works for the case $\rho _{0}^{B}\neq \rho _{1}^{B}$, thus it fills the gap
where the MLC proof left. Meanwhile, the MLC theorem concentrates on the
cheating of Alice. It does not exclude the existence of protocols which is
unconditionally secure against dishonest Bob only. On the other hand, our
result shows that Bob can always cheat in CSQBC regardless Alice is honest
or not.

%While there is no proof that the above features are sufficiently
%general to exhaust all possible CSQBC protocols, it is worth noting the fact
%that all existing CSQBC protocols \cite{qi150,qbc50,qbc78,qi197,qbc52,qbc89}
%are covered. To be precise, our security analysis here is restricted only to
%the CSQBC protocols having these features.

It will be interesting to study whether there can be CSQBC protocols without
the above four features. It seems that Kent's relativistic QBC \cite%
{qi44,qbc24,qbc51} and our recent proposals \cite{HeJPA,HeQIP} do not
satisfy feature (1), while the protocol in \cite{HePRA} does not have
feature (2), as elaborated in \cite{HeProof}. However, these works are aimed
to achieve the original QBC, instead of CSQBC. Also, \cite%
{HeJPA,HeQIP,HePRA,HeProof} have not gained wide recognition yet. Thus it is
still an open question whether it is possible to build non-relativistic
CSQBC protocols which are not limited by the above security bounds, without
relying on computational and experimental constraints.

%The purpose of our current result is not to put a limit on the power of CSQBC.
%Instead, we wish it could provide clues for building CSQBC that can exceed the
%security bound.

\section*{Methods}

\subsection*{Calculating Bob's cheating probability}

Consider the POVM $\{P_{0}^{\dag }P_{0},P_{1}^{\dag }P_{1}\}$ defined in Eq.
(\ref{projector2}). After Bob applies it on $\Psi $, there can be two
outcomes.

(I) The projection outcome is $P_{0}$. Then Bob takes $b^{\prime }=0$ as his
decoded result of Alice's committed bit $b$. With Eqs.
%(\ref{projector2}),
(\ref{projector2}) and (\ref{initial}) we yield%
\begin{equation}
P_{0}\left\vert \Phi \otimes \Psi \right\rangle _{ini}=\sqrt{\alpha }%
\sum\limits_{i}\lambda _{i}^{(0)}\left\vert f_{i}^{(0)}\right\rangle \otimes
\left\vert e_{i}^{(0)}\right\rangle .
\end{equation}%
Thus this case will occurs with the probability%
\begin{equation}
p_{I}=\alpha ,  \label{pI}
\end{equation}%
while the resultant state of $\Phi \otimes \Psi $ is%
\begin{equation}
\left\vert \Phi \otimes \Psi \right\rangle _{I}=\frac{1}{\sqrt{p_{I}}}%
P_{0}\left\vert \Phi \otimes \Psi \right\rangle _{ini}.
\end{equation}

As described in feature (3) of CSQBC, at the unveil phase Alice may require
Bob to return $\Psi $ and check whether it remains intact in its initial
state. The maximal probability for Alice to find out that Bob has already
projected $\left\vert \Phi \otimes \Psi \right\rangle _{ini}$\ into $%
\left\vert \Phi \otimes \Psi \right\rangle _{I}$\ is bounded by%
\begin{align}
\tilde{p}_{I}& =1-\left\vert _{I}\left\langle \Phi \otimes \Psi \right.
\left\vert \Phi \otimes \Psi \right\rangle _{ini}\right\vert ^{2}  \nonumber
\\
& =1-\frac{1}{p_{I}}\alpha ^{2}.
\end{align}%
Thus the total probability for (case (I) occurred) $AND$ (Alice failed to
detect Bob's cheating) is%
\begin{equation}
p_{I}(1-\tilde{p}_{I})=\alpha ^{2}.
\end{equation}

(II) The projection outcome is $P_{1}$. Then Bob takes $b^{\prime }=1$ as
his decoded result of Alice's $b$. Now%
\begin{equation}
P_{1}\left\vert \Phi \otimes \Psi \right\rangle _{ini}=\sqrt{\beta }%
\sum\limits_{i}\lambda _{i}^{(1)}\left\vert f_{i}^{(1)}\right\rangle \otimes
\left\vert e_{i}^{(1)}\right\rangle .
\end{equation}%
Obviously, this case will occurs with the probability%
\begin{equation}
p_{II}=1-p_{I}.
\end{equation}%
Meanwhile, the resultant state of $\Phi \otimes \Psi $ in this case is%
\begin{equation}
\left\vert \Phi \otimes \Psi \right\rangle _{II}=\frac{1}{\sqrt{p_{II}}}%
P_{1}\left\vert \Phi \otimes \Psi \right\rangle _{ini}.
\end{equation}

The maximal probability for Alice to find out that Bob has already projected
$\left\vert \Phi \otimes \Psi \right\rangle _{ini}$\ into $\left\vert \Phi
\otimes \Psi \right\rangle _{II}$\ is bounded by%
\begin{align}
\tilde{p}_{II}& =1-\left\vert _{II}\left\langle \Phi \otimes \Psi \right.
\left\vert \Phi \otimes \Psi \right\rangle _{ini}\right\vert ^{2}  \nonumber
\\
& =1-\frac{1}{p_{II}}\beta ^{2}.
\end{align}%
Thus the total probability for (case (II) occurred) $AND$ (Alice failed to
detect Bob's cheating) is%
\begin{equation}
p_{II}(1-\tilde{p}_{II})=\beta ^{2}.
\end{equation}

Taking both cases (I) and (II) into consideration, the overall probability
for Bob's cheating to pass Alice's detection successfully is%
\begin{eqnarray}
P_{B} &=&p_{I}(1-\tilde{p}_{I})+p_{II}(1-\tilde{p}_{II})=\alpha ^{2}+\beta
^{2}  \nonumber \\
&=&\frac{1}{2}+\frac{1}{2}(2\alpha -1)^{2}.  \label{pc*}
\end{eqnarray}

Meanwhile, since the projection outcome will either be $P_{0}$ or $P_{1}$\
with the probabilities $p_{I}$\ and $p_{II}=1-p_{I}$, respectively, Bob's $%
b^{\prime }$ will match Alice's $b$ with the probability $p_{I}$\ or $%
1-p_{I} $ too. Note that $h(1-p_{I})=h(p_{I})$.
%, where $h(p)\equiv -p\log _{2}p-(1-p)\log _{2}(1-p)$\ is the binary entropy function.
Thus the amount of mutual information that Bob obtains with this POVM is%
\begin{equation}
I_{m}=1-h(p_{I})=1-h(\alpha ).  \label{Im*}
\end{equation}

\subsection*{Bounding the cheating probabilities with trace distance}

Suppose that there are many states allowed to be chosen randomly for
committing $b=0$ in the protocol, each of which takes the form of Eq. (\ref%
{initial}),
% (if there is the system $E$, in Bob's view it can be treated as a
%part of Alice's system $\Phi $)
but with different values of the coefficients $\alpha $, $\beta $, $\lambda
_{i}^{(0)}$'s and $\lambda _{i}^{(1)}$'s. Bob applies the optimal POVM
% (i.e., taking $q=1$ in Eq. (\ref{projector}))
to decode $b$. Then Eq. (\ref{initial}) indicates that he can learn $b$
correctly with probability $\bar{\alpha}$, i.e., the average of $\alpha $.
Meanwhile, it is well-known that the maximal probability for discriminating
two density matrices $\rho _{0}^{B}$, $\rho _{1}^{B}$ is $(1+D(\rho
_{0}^{B},\rho _{1}^{B}))/2$.\ Therefore%
\begin{equation}
D(\rho _{0}^{B},\rho _{1}^{B})=2\bar{\alpha}-1.
\end{equation}%
Since Eq. (\ref{pc}) shows that Bob's average cheating probability for these
states is%
\begin{equation}
P_{B}=\frac{1+\overline{(2\alpha -1)^{2}}}{2}\geq \frac{1+(2\bar{\alpha}%
-1)^{2}}{2},  \label{37}
\end{equation}%
we have%
\begin{equation}
P_{B}\geq \frac{1+D(\rho _{0}^{B},\rho _{1}^{B})^{2}}{2}.  \label{pb lower M}
\end{equation}%
Similar discussion is also valid for the states for committing $b=1$, except
that $\alpha $ should be replace by $\beta =1-\alpha $. But Eq. (\ref{pb
lower M}) remains the same because Eq. (\ref{pc}) satisfies $P_{B}(1-\alpha
)=P_{B}(\alpha )$.

On the other hand, since \cite{qi366}%
\begin{equation}
F(\rho _{0}^{B},\rho _{1}^{B})\geq 1-D(\rho _{0}^{B},\rho _{1}^{B}),
\end{equation}%
from Eq. (\ref{pa}) we yield

\begin{equation}
P_{A}\geq 1-\frac{D(\rho _{0}^{B},\rho _{1}^{B})}{2}.  \label{pa lower E}
\end{equation}%
%
%
%
%
%
%We would like to replace $D(\rho _{0}^{EB},\rho _{1}^{EB})$ with $D(\rho
%_{0}^{B},\rho _{1}^{B})$\ so that the above equation can be studied together
%with Eq. (\ref{pb lower}). However, it is a common feature of trace distance
%that $D(\rho _{0}^{EB},\rho _{1}^{EB})\geq D(\rho _{0}^{B},\rho _{1}^{B})$,
%i.e., $1-D(\rho _{0}^{B},\rho _{1}^{B})/2\geq 1-D(\rho _{0}^{EB},\rho
%_{1}^{EB})/2$. This makes it hard to find the lower bound of $P_{A}$ using $%
%D(\rho _{0}^{B},\rho _{1}^{B})$, because in general it is unknown whether $%
%P_{A}$ or $1-D(\rho _{0}^{B},\rho _{1}^{B})/2$\ is larger.

\section*{Acknowledgements}

The work was supported in part by the NSF of China, the NSF of Guangdong
province, and the Foundation of Zhongshan University Advanced Research
Center.

\section*{Additional information}

\textbf{Competing financial interests:} The author declares no competing
financial interests.


\begin{thebibliography}{99}
\bibitem{qi75} Yao, A. C. C. Security of quantum protocols against coherent
measurements. In \textit{Proc. 26th Symposium on the Theory of Computing}.
New York: ACM, pp. 67. (1995).

\bibitem{qi139} Kilian, J. Founding crytpography on oblivious transfer. In
\textit{Proc. 1988 ACM Annual Symposium on Theory of Computing}. New York:
ACM, pp. 20. (1988).

\bibitem{qi24} Mayers, D. Unconditionally secure quantum bit commitment is
impossible. \textit{Phys. Rev. Lett.} \textbf{78}, 3414 (1997).

\bibitem{qi23} Lo, H. -K. \& Chau, H. F. Is quantum bit commitment really
possible? \textit{Phys. Rev. Lett.} \textbf{78}, 3410 (1997).

\bibitem{qi150} Hardy, L. \& Kent, A. Cheat sensitive quantum bit
commitment. \textit{Phys. Rev. Lett.} \textbf{92}, 157901 (2004).
%\textit{quant-ph/9911043v5}.

\bibitem{qbc50} Aharonov, D., Ta-Shma, A., Vazirani, U. V. \& Yao, A. C.
Quantum bit escrow. \textit{arXiv:quant-ph/0004017v1}. In \textit{Proc. 32nd
Annual Symposium on Theory of Computing}. New York: ACM, pp. 705. (2000).

\bibitem{qbc78} Jakoby, A., Liskiewicz, M. \& Madry, A. Using quantum
oblivious transfer to cheat sensitive quantum bit commitment. \textit{%
arXiv:quant-ph/0605150v1} (2006).

\bibitem{qi197} Buhrman, H., Christandl, M., Hayden, P., Lo, H. -K. \&
Wehner, S. Possibility, impossibility, and cheat sensitivity of quantum-bit
string commitment. \textit{Phys. Rev. A} \textbf{78,} 022316 (2008).
%\textit{quant-ph/0504078v2}.

\bibitem{qbc52} Shimizu, K., Fukasaka, H., Tamaki, K. \& Imoto, N.
Cheat-sensitive commitment of a classical bit coded in a block of m$\times $%
n round-trip qubits. \textit{Phys. Rev. A} \textbf{84,} 022308 (2011).

\bibitem{qbc89} Li, Y. -B., Wen, Q. -Y., Li, Z. -C., Qin, S. -J. \& Yang, Y.
-T. Cheat sensitive quantum bit commitment via pre- and post-selected
quantum states. \textit{Quant. Inf. Process.} \textbf{13}, 141 (2014).

\bibitem{qi73} Hughston, L. P., Jozsa, R. \& Wootters, W. K. A complete
classification of quantum ensembles having a given density matrix. \textit{%
Phys. Lett. A} \textbf{183}, 14 (1993).

\bibitem{qi366} Nielsen, M. A. \& Chuang, I. L. in \textit{Quantum computation and
quantum information}, Ch. 9.2, 404-416 (Cambridge, 2000).

\bibitem{HeComment} He, G. P. Comment on \textquotedblleft Cheat-sensitive
commitment of a classical bit coded in a block of m$\times $n round-trip
qubits\textquotedblright . \textit{Phys. Rev. A} \textbf{89}, 056301 (2014).

%\bibitem{qi581} Kent, A. Impossibility of unconditionally secure commitment
%of a certified classical bit. \textit{Phys. Rev. A} \textbf{61,} 042301
%(2000). %\textit{quant-ph/9910087v2}

\bibitem{qi365} Bennett, C. H. \& Brassard, G. Quantum cryptography: public
key distribution and coin tossing. In \textit{Proceedings of the IEEE
International Conference on Computers, Systems, and Signal Processing,} 175
(IEEE Press, New York, 1984).

\bibitem{qi442v3} Ishizaka, S. Is cheat sensitive quantum bit commitment
really possible? \textit{arXiv:quant-ph/0703099v3} (2007).

\bibitem{qi442} Ishizaka, S. Dilemma that cannot be resolved by biased
quantum coin flipping. \textit{Phys. Rev. Lett.} \textbf{100}, 070501
(2008). %\textit{quant-ph/0703099v5}.

\bibitem{qi44} Kent, A. Unconditionally secure bit commitment. \textit{Phys.
Rev. Lett.} \textbf{83}, 1447 (1999). %\textit{quant-ph/9810068v4}.

\bibitem{qbc24} Kent, A. Unconditionally secure bit commitment with flying
qudits. \textit{New J. Phys.} \textbf{13,} 113015 (2011).
%\textit{arXiv:1101.4620v4}.

\bibitem{qbc51} Kent, A. Unconditionally secure bit commitment by
transmitting measurement outcomes. \textit{Phys. Rev. Lett.} \textbf{109},
130501 (2012). %\textit{arXiv:1108.2879v2}.

\bibitem{HeJPA} He, G. P. Quantum key distribution based on orthogonal
states allows secure quantum bit commitment. \textit{J. Phys. A: Math. Theor.%
} \textbf{44}, 445305 (2011).

\bibitem{HeQIP} He, G. P. Simplified quantum bit commitment using single
photon nonlocality. \textit{Quantum Inf. Process.} \textbf{13}, 2195 (2014).

\bibitem{HePRA} He, G. P. Secure quantum bit commitment against empty
promises. \textit{Phys. Rev. A} \textbf{74}, 022332 (2006).

\bibitem{HeProof} He, G. P. Secure quantum bit commitment against empty
promises. II. The density matrix. \textit{arXiv:1307.7318} (2013).
\end{thebibliography}
\end{document}